\documentclass[12pt]{article}

\usepackage{graphicx}
\usepackage{epstopdf}
\newcommand{\beq}{\begin{equation}}
\newcommand{\eeq}{\end{equation}}
\newcommand{\beqa}{\begin{eqnarray}}
\newcommand{\eeqa}{\end{eqnarray}}
\newcommand{\beqar}{\begin{eqnarray*}}
\newcommand{\eeqar}{\end{eqnarray*}}

\begin{document}
\thispagestyle{empty}

\hfill{\sc UG-FT-257/09}

\vspace*{-2mm}
\hfill{\sc CAFPE-127/09}

\vspace{32pt}
\begin{center}

\textbf{\Large Black hole gas in the early universe\\
} 
\vspace{40pt}

M\'onica Borunda, Manuel Masip
\vspace{12pt}

\textit{
$^{1}$CAFPE and Departamento de F{\'\i}sica Te\'orica y del
Cosmos}\\ 
\textit{Universidad de Granada, E-18071 Granada, Spain}\\
\vspace{8pt}

\texttt{mborunda@ugr.es, masip@ugr.es}
\end{center}

\vspace{40pt}

\date{\today}

\begin{abstract}
We consider the early universe at temperatures close to
the fundamental scale of gravity ($M_D\ll M_{Planck}$) 
in models with extra dimensions.
At such temperatures a small fraction 
of particles will experience transplanckian collisions that may 
result in microscopic black holes (BHs). BHs colder than the 
environment will gain mass, and as they grow their temperature drops
further. We study the dynamics of a system (a {\it black hole gas})
defined by radiation at
a given temperature coupled to a distribution of BHs of different 
mass. Our analysis includes the production of BHs in photon-photon 
collisions, BH evaporation, the absorption of radiation, 
collisions of two BHs to give a larger one, and the effects of the
expansion. We show that the system may follow two different
generic paths depending on the initial temperature of the plasma.

\end{abstract}

                            

\newpage

\section{Introduction}

Large (ADD) \cite{ArkaniHamed:1998rs} 
or warped (RS) \cite{Randall:1999ee} extra dimensions open the possibility 
that the fundamental scale of gravity $M_D$ is much lower than 
$M_P\approx 10^{19}$ GeV.
This could imply that the {\it transplanckian}
regime \cite{Banks:1999gd} is at accesible energies.
Collisions in that regime are
very different from what we have experienced so far in
particle colliders:
due to its spin 2 the graviton becomes strongly coupled and
dominates at distances that increase with the center of mass
energy $\sqrt{s}$. 
In particular, at small impact parameters one expects that
gravity {\it bounds} the system and the two particles collapse
into a microscopic black hole (BH) of mass around $\sqrt{s}$.
Such BH would evaporate \cite{Hawking:1974rv} very fast 
(the time scale is 
set by $1/M_D$) into final states of high multiplicity,
a possible LHC signature that has been extensively discussed
in the literature \cite{Dimopoulos:2001hw}.

These collisions, however, may have also occurred in the early
universe if the temperature was ever close to $M_D$. 
Consider
a period of inflation produced by a field on the brane, 
followed by reheating into 4--dimensional (4-dim) species. 
If the 
reheating temperature is $T_{rh}\ge 0.1 M_D$, particles
in the tail of the Boltzmann distribution may
collide with enough energy to form BHs. Hot BHs will evaporate, 
but if a mini-BH is colder than the environment
it absorbs more than it emits \cite{Majumdar:2002mra}, 
growing and becoming 
colder, which in turn increases its absorption rate (see below).
The fact that heavier BHs live longer distinguishes them from 
massive particles or string excitations possibly  
produced at these temperatures, since the 
lifetime of the latter is inversely proportional
to their mass. As a consequence, one expects that 
BHs are a critical 
ingredient at temperatures near the fundamental scale.
Notice that these BHs are not the primordial 
ones formed from the gravitational collapse of density
fluctuations 
\cite{Carr:1974nx,Barrow:1991dn,Argyres:1998qn,Majumdar:2005ba} 
at lower temperature $T$.

Although BHs would also be present 
in a 4-dim universe at $T\sim M_P$, they 
are most relevant in TeV gravity models. The reason
is that the expansion of the universe is a long-distance process,
so its rate is dictated by $M_P$ (not $M_D$) also in models 
of TeV gravity. If the {\it bulk} is basically empty
\cite{ArkaniHamed:1998nn,Hannestad:2001nq,Starkman:2000dy} 
or {\it thin} (see below), 
then the expansion at $T\le M_D$ will be negligible in 
terms of the fundamental time scale $1/M_D$, and there is plenty
of time for collisions producing BHs to occur and for 
the BHs to
grow. In contrast, in a 4dim universe
the expansion rate at $T\approx M_P$ 
is of order $H^{-1}\approx 1/M_P$, the temperature
of the universe drops before BHs have grown, and once it 
goes below $T_{BH}$ they evaporate in the same 
time scale.

In this article we explore the implications of having
initial temperatures near the fundamental scale of
gravity. First we define a consistent set up for 
TeV gravity. Then we study the behaviour of a single BH inside
a thermal bath in an expanding universe. Finally 
we consider the generic case, a {\it black hole gas}, with
radiation coupled to a distribution of BHs of different 
mass. We find remarkable that the effect of these 
mini-BHs in the early universe has been
almost completely overlooked in the literature (the only
analysis that we have found is given by Conley and
Wizansky in \cite{Conley:2006jg}), 
although many authors have considered
temperatures close (and above) the Planck 
scale (see \cite{Brandenberger:1988aj} and references 
therein).

\section{Consistent TeV gravity models}

There are two generic frameworks that may imply  
unsuppressed gravity at $M_D\approx 1$ TeV (or
at any scale $M_D<M_P$). In the first one (ADD), $n$ 
compact dimensions of length $L<1$ mm introduce a large
number of KK excitations of the graviton. These
gravitons, of effective 4-dim mass proportional to
$m_c\equiv L^{-1}$,
couple very weakly ($\approx \sqrt{s}/M_P$) 
to ordinary matter. However, the large number
of effective gravitons $\approx (\sqrt{s}/m_c)^n$ involved
in a collision gives an amplitude of order one, 
\beq
{s\over M_P^2}\times \left( {\sqrt{s}\over m_c}\right)^n
\approx 1\;,
\eeq
at $\sqrt{s}=M_D$ if 
\beq
m_c= M_D \left( {M_D\over M_P}\right)^{2/n}\;.
\label{lmd}
\eeq
In contrast, in the second scenario (RS) the KK excitations
of the graviton have unsuppressed couplings to matter
but large masses, right below $M_D$, so a few KK modes
suffice to define an order one gravitational
interaction at that scale.

The ADD set up has the basic cosmological problem pointed out
in \cite{Hannestad:2001nq}. Essentially, at temperatures
$T\gg m_c$ KK gravitons 
will be abundantly produced in annihilations
of brane particles, due to
the large multiplicity of final states. If the initial temperature
is large these gravitons will change the expansion
rate at the time of primordial nucleosynthesis. Even if the
initial temperature is as low as 1 MeV, their late decay
will distort the diffuse gamma ray background in an unacceptable
way. Obviously, an initial temperature close to $M_D$ 
would bring too many massive gravitons.

One solution would be to consider RS models of TeV gravity. 
There, at $T\approx M_D$ bulk and brane species have similar 
abundances, but massive gravitons will decay  
fast once 
$T<m_c\approx 0.1 M_D$, returning all the energy to
the brane and defining acceptable 4-dim cosmological models.

Even within the ADD framework, however, 
we can consider {\it hybrid}
models where the connection between $m_c$ and $M_D$ is {\it not}
the one given in Eq.~(\ref{lmd}). This can be obtained, for
example, 
with a warp factor \cite{Giudice:2004mg}.
The effect would be to {\it push} 
the KK modes towards the 4dim brane, reducing the
effective compact volume to $V\approx (1/m_c)^n$ 
while increasing their coupling to matter,
\beq
{s\over M_P^2}\rightarrow {s\over M_D^2}
\left( {m_c\over M_D} \right)^n\;.
\eeq
In this way, a smaller number of KK modes will imply an
order one gravitational interaction at the same scale 
$\sqrt{s}=M_D$. 
If the {\it free} parameter $m_c$ takes the value in
Eq.~(\ref{lmd}) we recover ADD, whereas for $m_c$ approaching
$M_D$ we obtain RS. At distances below 
$1/m_c$ gravity would be higher dimensional (similar to ADD)
whereas at larger distances it becomes 4-dimensional (like
in the usual RS scenario).

In this framework the KK modes of the graviton are not 
produced at temperatures $T<m_c$. Therefore, $m_c>10$ MeV
avoids  
astrophysical bounds \cite{Hannestad:2003yd} for 
any number $n$ of extra dimensions \cite{Giudice:2004mg}
and $M_D=1$ TeV. 
On the other hand, as these KK gravitons
have stronger couplings
to matter, they can decay much faster than in the usual ADD
model or decouple at temperatures below their mass. By
changing $m_c$ (with $M_D$ fixed) 
it seems easy to obtain models with no gravitons
at the time of primordial nucleosynthesis that are consistent 
with all cosmological observations.

\section{Single black hole in a thermal bath}

Let us consider the hybrid framework described in the
previous section with $n$ extra dimensions,  
a fundamental scale $M_D$, and an independent 
compactification
mass $m_c\ge 1$ GeV. Gravity at energies above $M_D$ 
and distances 
smaller than $1/m_c$ is strongly coupled and 
higher dimensional (just like in ADD), so the radius
and temperature of a BH of mass $M$ are
\beq
r_H = {a_n\over M_D}\; \left( 
{M\over M_D} \right)^{1/(n+1)}\;;\;\;\;
T_{BH}={n+1\over 4\pi r_H}\;,
\label{rh}
\eeq
with 
\beq
a_n=\left( {2^n\pi^{(n-3)/2} 
\Gamma\left({n+3\over 2}\right)\over n+2} \right)^{1\over n+1}\;.
\eeq
Once a BH reaches a radius $r_H= 1/m_c$ and fills up the 
whole compact space its (4-dim) size 
will not keep growing, since
all the KK gravitons but the zero mode provide 
short distance interactions. The radius should start growing
significantly only when the usual 4-dim horizon (produced
by the massless graviton) is of order $1/m_c$, {\it i.e.},
for BH masses $M\approx M_P^2/ m_c$. Above this threshold 
BHs are basically 4-dim.

We will assume that the particles in the brane and the bulk 
(with $g_*$ and $g_b$ degrees of freedom, respectively) 
are initially in thermal 
equilibrium at a temperature $T=T_0$. Notice that 
reheating in just the brane after a period of inflation 
would {\it not} justify an empty bulk, as in a time of
order $M_D^2/T^3$ most of the energy would escape into KK modes.
Here, however, the bulk may be much thinner and {\it emptier}
than in the usual ADD model, implying a slow expansion 
in terms of the fundamental time $1/M_D$ (see below).

It is easy to see that if the plasma temperature $T$ and
$1/r_{H}$ are larger than $m_c$ a BH at rest in the brane will 
change its mass according to \cite{Conley:2006jg}
\beqa
{{\rm d} M\over {\rm d} t} &\approx & 
\sigma_4 A_4 (T^4-T_{BH}^4)+\sigma_{4+n} A_{4+n} 
(T^{4+n}-T_{BH}^{4+n})\cr
&\approx & {\pi \over 480} \left({n+1\over T_{BH}}\right)^2
\left( g_* \left( T^4-T_{BH}^4 \right) + g_b c_n 
\left( {T^n\over T_{BH}^n}\; T^4 - T_{BH}^4\right)\right)
\;,
\label{dmdt}
\eeqa
where $\sigma_4=g_* \pi^2/120$, $c_n$ is an order 1 coefficient that
depends on the geometry of the compact space, and we have neglected 
gray-body factors \cite{Page:1976df}.
The expression above assumes a thermalized photon (in 
the brane) and graviton (in the bulk) plasma, 
so it requires that 
changes occur {\it slowly}. In particular, if the BH grows 
very fast one has to make sure that the gain in mass is always
smaller than the total energy of the plasma in
causal contact with the hole:
\beq
{{\rm d} M\over {\rm d} t} < {4 \pi^3 t^2\over 30}\; 
T ^4 \left( g_* + g_b c_n {T^n\over m_c^n} \right)\;,
\eeq
where we have assumed $t>1/m_c$.

As the BH grows it enters a new phase when its mass reaches 
\beq
M_1\approx M_D \left( {M_D \over a_n m_c} 
\right)^{n+1}\;, 
\eeq
which corresponds to a 
radius $r_H= 1/m_c$ filling up the whole extra volume.
For $M>M_1$ the BH keeps gaining mass as far as the plasma 
temperature $T$ is above $T_{BH}\approx m_c$.
However, as explained above, KK fields do not reach 
distances beyond 
$r_H\approx 1/m_c$, so the BH radius (and its temperature) will
stay basically constant.

Finally, masses above $M_2\approx M_p^2/m_c$ turn the BH 
into a purely 4 dimensional object: its radius 
\beq
r_{H} = {2M\over M_P^2} 
\eeq
grows larger than $1/r_c$ and the BH becomes 
too cold to emit massive gravitons. In this regime, 
if $T$ is larger than $m_c$ 
the BH changes its mass according to 
\beq
{{\rm d} M\over {\rm d} t} \approx 
{\pi \over 480}
\left( g_* + g_b c_n {T^n\over m_c^n} \right) 
{T^4\over T_{BH}^2}\;,
\label{dmdt2}
\eeq
whereas at lower plasma temperature it goes as
\beq
{{\rm d} M\over {\rm d} t} \approx 
{\pi \over 480}
g_* \left( 
{T^4\over T_{BH}^2}-T_{BH}^2\right) \;.
\label{dmdt3}
\eeq

Eq.(\ref{dmdt}) implies that lighter (hotter) BHs 
evaporate with an approximate lifetime
\beq
\tau\approx {1\over M_D} \left({M\over M_D}\right)^{3+n\over 1+n}\;, 
\eeq
and that BHs colder than the 
plasma will gain
mass: heat flows from the hot plasma to the cold
BH, but the effect is to cool the BH further and increase
$T-T_{BH}$. This is, indeed, a very
peculiar two-component thermodynamical system.

On the other hand, the expansion of the universe is also 
affected by
the presence of bulk species. We will assume that the extra
dimensions are {\it frozen} (do not expand) 
and will integrate the matter content
in the bulk. The large values of $m_c$ that we will 
consider imply large enough couplings with brane photons,
so that the (equilibrium) 
abundance of KK modes of mass larger than the plasma 
temperature will be
negligible. Therefore, at $T>m_c$ we have a universe with a
radiation density
\beq
\rho_{rad} \approx  {\pi^2\over 30}\; 
T ^4 \left( g_* + g_b c_n {T^n\over m_c^n} \right)\;, 
\label{rho}
\eeq
and an expansion rate
\beq
\frac{\dot R^2}{R^2} =\frac{8\pi G}{3} \rho_{rad}\;,
\eeq
{\it i.e.}, $\rho_{rad}=\rho_{rad0}(R_0/R)^4$.
Notice that the second term in Eq.(\ref{rho}) will 
slow down the change in the plasma temperature $T(t)$ due to
the expansion.
In particular, if the bulk energy dominates then 
$T\propto t^{2/(4+n)}$ for times larger than a Hubble time.
At temperatures below $m_c$ this term vanishes exponentially 
and all the bulk energy is transferred to the brane.

To ilustrate the orders of magnitude involved, let us discuss
a toy model with 
$M_D=1$ TeV, $n=1$ ($c_1=0.3$), and just photons ($g_*=2$) and
gravitons ($g_b=5$) at $T_0=100$ GeV. 
A BH of mass $M<M_{crit}=12$ TeV  would be hotter than the
environment, and it would evaporate in a time of order
$\tau\approx 10^{-3}$ GeV$^{-1} = 6.5\times 10^{-28}$ s.
The Hubble time of a universe at this temperature is
\beq
H^{-1}={R\over \dot R}=1.4\times 10^{14}\; {\rm GeV^{-1}} 
= 9.2\times 10^{-9}\; {\rm s}\;.
\eeq
A BH of
initial mass $M_0=100$ TeV will have a 
starting temperature of 8.7 GeV. In a time of order 17 GeV$^{-1}$
its mass is already around $4.7\times 10^7$ GeV and 
its radius as large as the
size of the extra dimension ($0.1$ GeV$^{-1}$). 
Then the BH keeps growing at approximately constant rate, since
its size and temperature ($T_{BH}\approx 1.6$ GeV)
change little with the mass. In a Hubble time 
the BH reaches a mass $M\approx 5.7\times 10^{19}$ GeV. 
At later times the expansion cools the
plasma, which {\it slows} the growth of the BH. Its maximum mass, 
$M\approx 2\times 10^{21}$ GeV, is achieved when 
$T = T_{BH}$
at times of order $t\approx 10^{18}$ GeV$^{-1}$. 
Finally, the BH will evaporate
after $\tau\approx 10^{22}$ GeV$^{-1}\approx 1$ s.

In Fig.~1 we plot this case together with the mass evolution
\begin{figure}
\begin{center}
\includegraphics[width=0.5\linewidth]{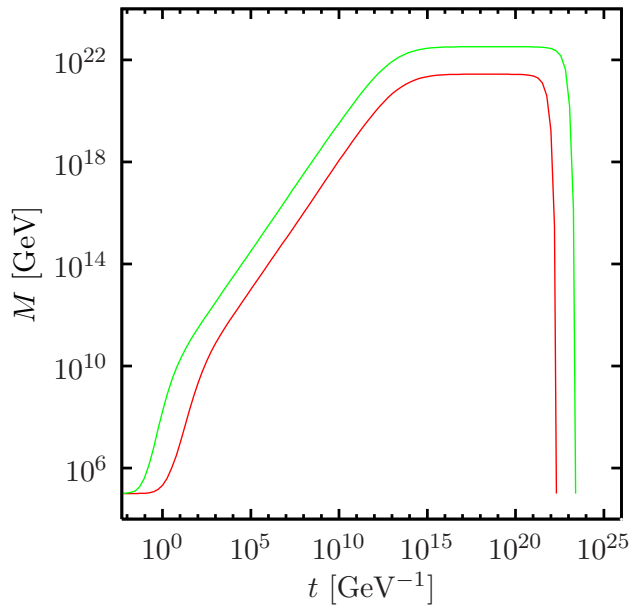}
\end{center}
\caption{Evolution of a single BH of mass $M_0=100$ TeV 
for $M_D=1$ TeV and $n=1$ in an expanding universe
at $T_0=100$ (lower line) and 
200 GeV (upper line).
\label{fig1}}
\end{figure}
for a larger initial temperature, $T_0=200$ GeV. 
In this second case the expansion rate is
faster (the Hubble time $H^{-1}\approx 2.6\times 10^{13}$ GeV$^{-1}$
is shorter), but the 
higher radiation density provides larger BH masses. 
Lower (higher) values of $m_c$ ($M_D$) 
would also imply larger BH masses.

In none of the two cases described above the 
BH becomes purely 4 dimensional, {\it i.e.,} with a
mass $M>10^{37}$ GeV. These large values of $M$ can be obtained
increasing the number $n$ of extra dimensions and/or
the ratio $M_D/m_c$.
A BH radius larger than $1/m_c$, 
however,  
would not stop its growth, on the 
contrary, it would enhance the absorption rate 
of radiation by the BH. 
What stops the growth of a BH in any TeV-gravity framework is just
the drop in the temperature of the plasma due to the expansion.

It is also important 
to emphasize that the qualitative features of the
process described above do not
depend much on the details of the compactification 
(the length or the shape) of the extra dimensions.
They just depend on
the fact that the fundamental scale of gravity $M_D$ 
is low and that
the initial temperature of the 
radiation is close to it. In particular, the sequence
of events would be similar in RS models, although there 
$m_c$ is close to $M_D$ 
and the BHs are smaller ($r_H\approx 1/m_c$, and
the absortion rate grows with the BH area). 

Finally, notice that the usual cosmological constraints 
\cite{Hannestad:2001nq,Starkman:2000dy} on the initial 
(reheating) temperature of the early universe 
or the astrophysical bounds on $M_D$ 
are actually a probe of $1/L=m_c$, so they are avoided
if $m_c$ is above 1 GeV.

\section{Dynamics of a black hole gas}

Let us now deduce the equations that describe the production
and evolution of 
BHs inside a thermal bath of temperature $T$.
If $T<M_D < M$ the BHs will be produced in collisions
between radiation in the high-energy tail of the Boltzmann 
distribution 
$f_\gamma (\vec k)$. Once formed these BHs will be 
non-relativistic,
with a kinetic energy $K\approx T$ negligible versus the
mass $M$ and a velocity $v\approx \sqrt{2 T/M}$. 
We will then assume that the two components
of the system are well described by the temperature $T(t)$ 
of the plasma and
the distribution $f(M,t)$ expressing the number of BHs of mass
$M$ per unit mass and volume at time $t$.
The total energy density is 
\beq 
\rho(t)=
\rho_{rad}(t)+\rho_{BH}(t)\approx 
{\pi^2\over 30}\; 
T ^4 \left( g_* + g_b c_n {T^n\over m_c^n} \right)
+\int {\rm d}M\; M \;f(M,t)\;.
\eeq

We identify the following processes changing the number density 
of BHs of mass $M$.

\begin{itemize}

\item BH production in photon-photon collisions.
Photons will collide with a cross section  
$\sigma\approx \pi r_H^2$ to form a BH of mass $M\approx \sqrt{s}$:
\beq
\left( {\partial f(M,t)\over \partial t} \right)_{\gamma\gamma\rightarrow
M}=\int\frac{{\rm d}^3k_1}{(2\pi )^3}\frac{{\rm d}^3k_2}{(2\pi )^3} 
\;f_\gamma(\vec k_1) f_\gamma(\vec k_2)\;\sigma ( M )
\;|\vec v_1-\vec v_2|\; \delta(\sqrt{(k_1^\mu +k_2^\mu)^2}-M)\;,
\eeq
with $v_i=1$.
If $T \ll M$ this expression can be approximated in terms
of a modified Bessel function, 
\beq
\left( {\partial  f(M,t)\over \partial t} \right)_{\gamma\gamma\rightarrow
M}\approx {g_*^2 a_n^2\over 16 \pi^3}\; T M^2
\left( {M\over M_D}
\right)^{4+2n\over 1+n} K_1(M/T) 
\;.
\eeq
To simplify our analysis we will neglect BH production in
collisions of bulk particles. Although 
KK modes dominate the energy
density, their cross section to form a BH is smaller 
(their wave function is {\it diluted}
within the bulk), so this would be
an order one contribution.

\item The collision of two BHs, of mass $M_1$ and
$M_2$, to form a BH of mass $M=M_1+M_2$:
\beq
\left( {\partial  f(M,t)\over \partial t} \right)_{M_1 M_2\rightarrow
M}= \int {\rm d}M_1 {\rm d}M_2 \; f(M_1,t)f(M_2,t) 
\;\sigma (M_1 ,M_2) \;v_{12}\;
\delta (M_1 +M_2 - M)\;,
\eeq
where the BH velocity is $v_i=\sqrt{2T/M_i}$ and 
$v_{12}= \langle |\vec v_1-\vec v_2| \rangle$.
We will take a BH--BH cross section of 
\beq
\sigma (M_1 ,M_2)=\pi (r_{H1}+r_{H2})^2\;.
\eeq
Notice also that for a minimum BH mass of $M_D$, 
this contribution to $f(M,t)$ is nonzero only if $M\ge 2 M_D$.

\item A BH of mass $M$ may collide with
any other BH, which would reduce $f(M,t)$:
\beq
\left( {\partial  f(M,t)\over \partial t} \right)_{M M_1\rightarrow
M_2} = - \int {\rm d}M_1\;  f(M,t)f(M_1,t) 
\;\sigma (M ,M_1) \;v_{01}\;,
\eeq

\item We can describe the absorption
and emision of radiation using 
${\rm d}M/{\rm d}t$ in (\ref{dmdt}). The BHs of mass $M$ will have
a mass $M+{\rm d}M$ at $t+{\rm d}t$, {\it i.e.},
\beq
f(M,t)=f(M+{\rm d}M,t+{\rm d}t)\;.
\eeq
This implies
\beq
\left( {\partial  f(M,t)\over \partial t} \right)_{abs/em} 
=  {\partial  f(M,t)\over \partial M} \; {{\rm d}M\over {\rm d}t}\;.
\eeq

\item Finally, we have to add the effect of the expansion. The
4-dim scale factor grows according to 
\beq
\frac{\dot R^2}{R^2} =\frac{8\pi G}{3} \left(\rho_{rad}+
\rho_{BH}\right)\;.
\eeq
This dilutes the number of BHs at a rate 
\beq
\left( {\partial f(M,t)\over \partial t} \right)_{exp}=
-3 \; f(M,t) {\dot R\over R}\;.
\eeq

\end{itemize}
The total change in $f(M,t)$ per unit time will result from the 
addition of the 5 contributions above. This  fixes $\dot \rho_{BH}$:
\beq
\dot \rho_{BH} = \int {\rm d}M\; M \;
{\partial f(M,t)\over \partial t}\;.
\eeq
On the other hand, to obtain the change in $T$ (or
$\rho_{rad}(T)$) we impose 
energy conservation, 
\beq
{\rm d}(\rho_{rad} R^3) + 
{\rm d}(\rho_{BH}R^3)=-\frac 13 \rho_{rad} \;{\rm d} R^3\;,
\label{ec}
\eeq
where we have neglected the pressure of the BHs (they behave
like non-relativistic matter).
The equation above implies
\beq
\dot \rho_{rad} = - 4 \rho_{rad} {\dot R \over R}
-\dot \rho_{BH} - 3 \rho_{BH} {\dot R \over R}\;.
\eeq

Notice that 
if the radiation and the BHs where decoupled, then one would have
${\rm d}(\rho_{BH}R^3)=0$, {\it i.e.},
$\dot \rho_{rad} = - 4 \rho_{rad} (\dot R/ R)$. 
This, however, is not the case since there is energy exchange between
the two components.
In particular, a variation in 
$\rho_{rad}$ changes $T$, implying
a change in the absorption/emission rate of the BHs and in $\rho_{BH}$.
It is easy to see that
\beq
{\rm d}(\rho_{BH}R^3)=R^3 \; \alpha \; {\rm d}\rho_{rad}\; \;,
\eeq
where we define $\alpha$ as 
\beq
\alpha \equiv \left( {\partial  \rho_{BH}\over 
\partial \rho_{rad}} \right)_{R=cons.}\;.
\eeq
Substituting this equation in (\ref{ec}) we obtain 
\beq
R^3\left( 1+\alpha \right) {\rm d}\rho_{rad}=
-{4\over 3}\;\rho_{rad}\;  {\rm d}R^3
\label{egamma}
\eeq

The equation above expresses that BH evaporation can slow down
the cooling of the radiation due to the expansion: as the universe
expands, $T$ drops, this leaves some BHs hotter than the 
plasma, so they evaporate and reheat 
the environment. Taking 
an interval where $\alpha$ 
is constant, eq.~(\ref{egamma}) can be integrated to
\beq
\rho_{rad} \approx \rho_{{rad} 0} 
\left( {R_0\over R}\right)^{4\over 1+\alpha}\;.
\eeq
If $\alpha\gg 1$ BH evaporation would {\it stop} the change in
the density and the temperature of the radiation due to the
expansion.

\section{Black-hole dominated plasma}

\begin{figure}
\begin{center}
\includegraphics[width=0.5\linewidth]{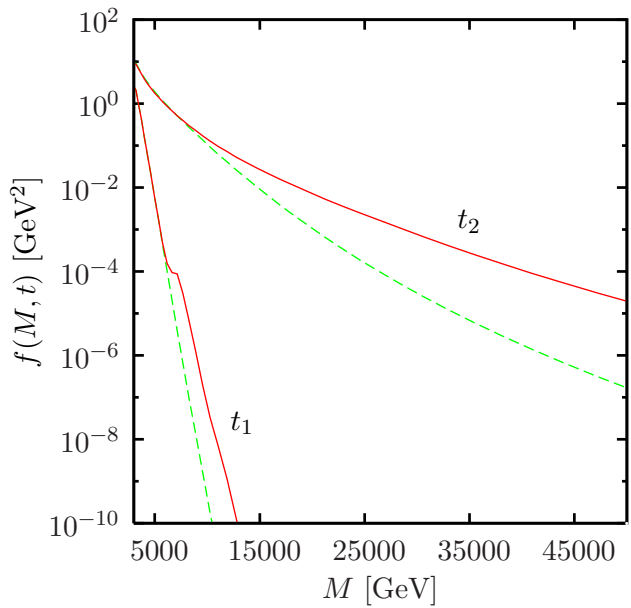}
\end{center}
\caption{Distribution $f(M,t)$ for 
$t_i=0.1,1$ GeV$^{-1}$ and $T_0=200$ GeV. Dashes describe the
evolution neglecting BH--BH collisions.
\label{fig2}}
\end{figure}
We will now apply these equations to an initial configuration with
only radiation (no BHs) at a given temperature $T(0)=T_0$. We find 
two generic scenarios
depending on the value of $T_0$. Values closer to $M_D$ produce a 
larger number of BHs, that grow and absorb all the radiation 
before a Hubble time. Lower values of $T_0$ imply a smaller number
of BHs, that grow
at basically constant temperature up to times of order $H^{-1}$. 
In this section we will focus on the first case.
We will describe the sequence of events using the toy model 
with $n=1$, $g_*=2$, $g_b=5$, 
$m_c=10$ GeV, $M_D=1$ TeV and an initial temperature of 
$T_0=200$ GeV. The critical BH mass (corresponding to $T_{BH}=T_0$) 
is $M_{crit}=3.0$ TeV. In Fig.~2 we plot the distribution
$f(M,t)$ for two values of $t$. We have included the distributions
with and without the effect of collisions of two BHs to form a
larger one.

We can distinguish four phases
in the evolution of the BH gas.

\begin{enumerate}

\item In a first phase BH's of $M>M_{crit}$ are 
produced at constant rate (see Fig.~3)
\begin{figure}
\begin{center}
\begin{tabular}{cc}
\includegraphics[width=0.45\linewidth]{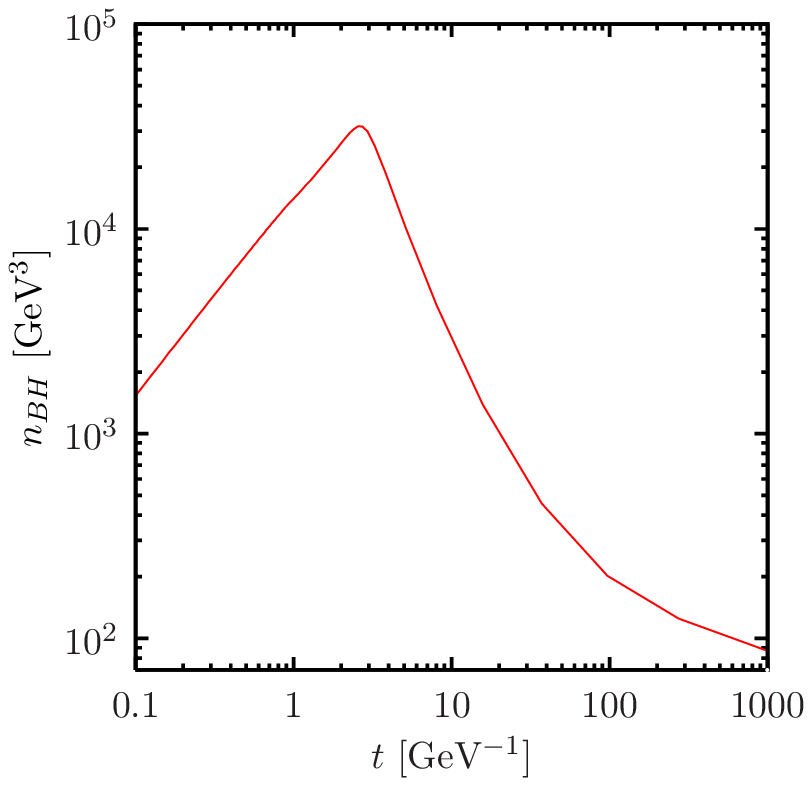} &
\includegraphics[width=0.45\linewidth]{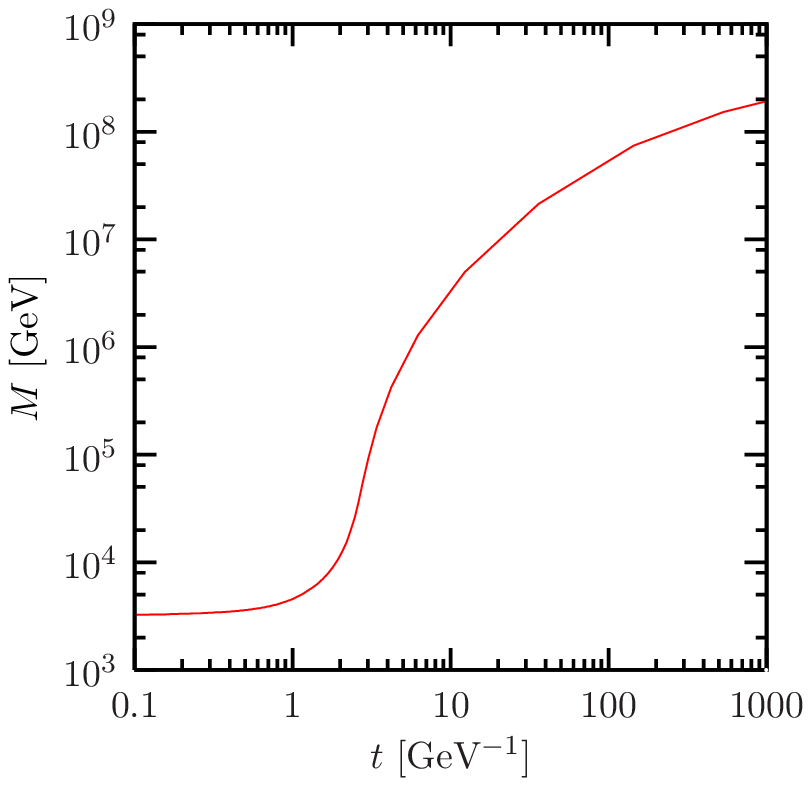}
\end{tabular}
\end{center}
\caption{Number of BHs per unit volume and average
BH mass as a function of time for $T_0=200$ GeV.
\label{fig3}}
\end{figure}
and absorb radiation of temperature $T\approx T_0$.
As the number of BHs grows (see Fig.~2), collisions between
two BHs to produce a larger one become important. This reduces the
number of BHs and increses their average mass (in Fig.~3). 
At times around $t\approx 2$ GeV$^{-1}$ 
BHs start dominating the energy density and the temperature
of the radiation drops (see Fig.~4). 

\item The drop in $T$ 
stops the production of BHs in 
photon-photon collisions. In addition, the lighter BHs become
hotter than the plasma, so they evaporate and feed $\rho_{rad}$. 
The evaporation reduces the number of BHs per unit volume, 
but the average BH mass grows: there is a
continuous transfer of energy from the radiation 
and from the lighter (evaporating) BHs to the larger BHs of 
lower temperature. Once the BHs get a mass around 
$5\times 10^7$ GeV their radius stops growing.

\item At $t\approx 10^4$ GeV$^{-1}$ the temperature of 
the radiation and of the BHs is similar, around 
$1.6$ GeV. The (slow) mass
growth of the heavier BHs is compensated by the decay of
the lighter ones, with the radiation temperature
basically constant. In this phase the 
energy density $\rho=\rho_{rad}+\rho_{BH}$ is 
matter (BH) dominated.

\item
At times of order $H^{-1}\approx 2.6\times 10^{13}$ GeV$^{-1}$ 
the expansion cools the radiation. The BHs, of mass around
$10^9$ GeV, decay fast ($\tau\approx 4\times 10^9$ GeV$^{-1}$) 
and the universe becomes radiation dominated. The lightest KK modes
also decay fast ($\tau_{KK}\approx M_D^{2+n}/m_c^{3+n}\approx
10^5$ GeV$^{-1}$), so only 4dim photons survive below
$T\approx 1$ GeV.
\begin{figure}
\begin{center}
\includegraphics[width=0.45\linewidth]{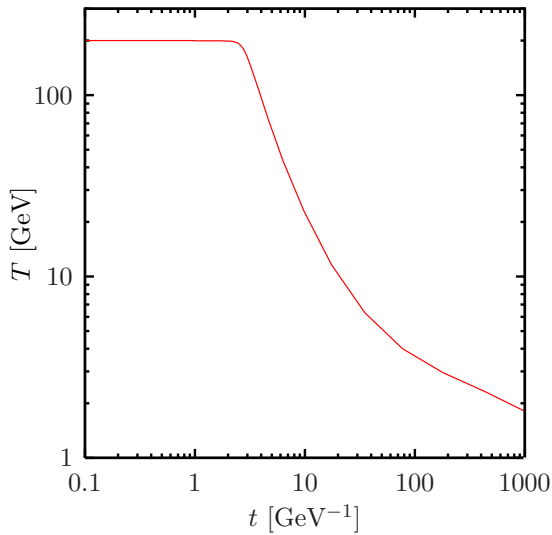} 
\caption{Temperature $T$ of the radiation as a function of 
$t$. At larger values of $t$ (up to a Hubble time 
$\approx 10^{13}$ GeV$^{-1}$) 
$T\approx 1.6$ GeV.\label{fig4}}
\end{center}
\end{figure}

\end{enumerate}

Two remarks are here in order. First, this is a toy model with only
photons, gravitons, and BHs. In a more complete set up one should 
include baryons at $T$ below 0.1 GeV 
(two types of matter, baryons and BHs, could coexist in models 
with lower values of $m_c/M_D$). Second, this generic high $T$ case,
with $M_D>5$ TeV to avoid bounds from colliders
and the inclusion of all the light standard model species (not
just photons) at each temperature, could define a realistic 
set up. In these hybrid models it seems {\it natural} to obtain 
a plasma dominated by the standard particles 
at $T\approx m_c \gg \Lambda_{QCD}$. The predictions for 
primordial nucleosynthesis
would then be consistent with observations.

\section{Radiation dominated plasma}

Let us now discuss the scenario with a low initial 
temperature and $\rho_{BH}\ll \rho_{rad}$ at any $t$.
We take $n=1$, $g_*=2$, $g_b=5$, 
$m_c=10$ GeV, $M_D=1$ TeV and 
$T_0=100$ GeV. The critical BH mass (corresponding to
$T_{BH}=T_0$) is $M_{crit}=12$ TeV, higher than in the previous 
case. As a consequence, the production rate of BHs colder than
the plasma is much smaller. We can separate three phases in 
the evolution of this BH gas.

\begin{enumerate}

\item At times below $H^{-1}=1.4\times 10^{14}$ GeV$^{-1}$ 
BHs are produced at constant
rate in photon-photon collisions, 
with $T\approx T_0$.
The number of BHs is so small that BH collisions can
be neglected. All these BHs grow like the one 
discussed in Section 2. 

\item When the expansion cools the photons, BH production
drops exponentially, whereas BH growth 
slows down. In Fig.~5 we plot the BH distribution $f(M,t)$
after one and four Hubble times.
The BHs reach a maximum mass
around $M=10^{21}\;{\rm GeV}$ 
(there is just a 10\% mass difference between
99\% of the BHs) and
$T_{BH}\approx 1.6$ GeV.
This universe is always radiation dominated. 
At $t\approx 10^{18}$ GeV$^{-1}$
the photon gas becomes colder than the BHs.
\begin{figure}
\begin{center}
\includegraphics[width=0.5\linewidth]{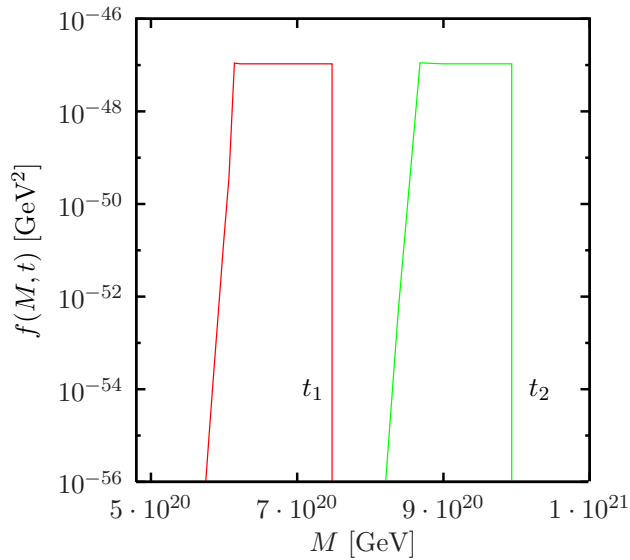}
\end{center}
\caption{Distribution $f(M,t)$ for 
$t_1=H^{-1}=1.4\times 10^{14}$ GeV$^{-1}$ and
 $t_2=4 H^{-1}$ with $T_0=100$ GeV.
\label{fig5}}
\end{figure}

\item In the last phase all the BHs evaporate
in a time scale $\tau\approx 10^{22}$ GeV$^{-1}$. 

\end{enumerate}

This generic scenario is in principle consistent with primordial
nucleosynthesis, since BHs are just a small fraction of the total
energy density and do not change the expansion rate. In the particular
case that we have discussed they decay when the plasma temperature
is $T\approx 0.01$ GeV, but increasing the ratio $M_D/m_c$ one can
obtain BHs that become 4 dimensional and with a much longer 
lifetime. Their late decay 
could introduce distortions in the diffuse gamma ray 
background. In addition, in a more complete set up 
including baryons and structure formation they might work 
as {\it seeds} for macroscopic (primordial) BHs.

\section{Summary and discussion}

A transplanckian regime at 
accessible energies would have new and {\it peculiar}
implications in collider physics and cosmology.
What makes this regime special is that as the energy
grows, softer physics dominate. For example,
the production of a {\it regular} heavy particle at the 
LHC would provide an event with very energetic 
(hard)
jets from its decay. In contrast, the production
of a mini-BH would be seen as a high multiplicity event,
with dozens of less energetic (soft) jets.
Analogously, the production of massive particles
or mini-BHs in the early universe would have very
different consequences. The heavier the elementary 
particle, the
shorter its lifetime, which tends to decouple these 
particles from low temperatures. For mini-BHs is 
just the opposite, heavier BHs are colder and
live longer.
In addition, if an approximate symmetry makes
a massive relic long lived, its late decay will 
produce very energetic
particles. BHs would imply a much
softer spectrum of secondaries \cite{Draggiotis:2008jz}, 
with different cosmological consequences.

In this paper we have analyzed the dynamics of a 
two-component gas with photons and BHs in an
expanding universe. The system is characterized
by the temperature $T(t)$ of the radiation
and the distribution $f(M,t)$ of BHs. Our equations
take into account BH production, absorption of photons,
BH evaporation, and collisions of two
BHs. We have discussed the two generic scenarios 
that may result from an initial temperature close
to the fundamental scale of gravity $M_D$. 
In the first scenario BHs {\it empty} of radiation
the universe and dominate $\rho$ 
before a Hubble time, whereas in the second case
there are few BHs that grow at constant $T$
up to $t\approx H^{-1}$.

We think that the work presented here is a 
necessary first step in the search for observable
effects from these BHs. It could also lead us to a
better understanding of the early universe at the
highest temperatures.

\section*{Acknowledgments}
We would like to thank Mar Bastero, Thomas Hahn and
Iacopo Mastromatteo for useful discussions.
This work has been supported by MEC of Spain 
(FIS2007-63364 and FPA2006-05294) and by 
Junta de Andaluc\'\i a (FQM-101 and FQM-437).
MB acknowledges a Juan de la Cierva fellowship from MEC of Spain.

\end{document}